\documentclass[epjCONF]{svjour}
\usepackage{graphics}
\usepackage[varg]{txfonts} 
\usepackage[latin1]{inputenc}
\session-title{NSRT12}
\begin{document}
\title{
A beyond-mean-field example with zero--range effective interactions in infinite nuclear matter}
\author{K. Moghrabi\inst{1}\fnmsep\thanks{\email{moghrabi@ipno.in2p3.fr}} \and M. Grasso\inst{1}
 \and X. Roca-Maza\inst{2}
  \and G. Col\`o\inst{2,3}
   \and N. Van Giai\inst{1}
}
\institute{Institut de Physique Nucl\'eaire, Universit\'e Paris-Sud, IN2P3-CNRS, F -91406 Orsay Cedex, France \and INFN, Sezione di Milano, Via Celoria 16, 20133 Milano, Italy \and Dipartimento di Fisica, Universit\`a degli Studi di Milano, 
Via Celoria 16, 20133 Milano, Italy
}
\abstract{
Zero--range effective interactions are commonly used in nuclear physics to describe a many-body system in the mean-field framework. 
If they are employed in beyond-mean-field models, an artificial ultraviolet divergence is generated by the zero-range of the interaction.
We analyze this problem in symmetric nuclear matter with the $t_0-t_3$ Skyrme model. In this case, the second-order energy correction diverges linearly with the momentum cutoff $\Lambda$. After that, we extend the work to the case of nuclear matter with the full Skyrme interaction. A strong divergence ($\sim\Lambda^5$) related to the velocity-dependent terms of the interaction is obtained. Moreover, a global fit can be simultaneously performed for both symmetric and nuclear matter with different neutron-to-proton ratios. These results pave the way for applications to finite nuclei in the framework of beyond mean-field theories.
} 
\maketitle
\section{Introduction}
\label{intro}
Contact interactions are reasonable approximations for the real finite-range forces that can be employed in cases where the interaction range is much smaller than the typical length-scale for the inter-particle distance. The main advantage of using zero--range interactions when treating many-body systems is that the equations to handle are usually simplified in this case. This is why contact interactions are extensively employed in many-body physics. Two examples of commonly adopted contact interactions are the zero-range effective Skyrme forces which are used in nuclear physics \cite{skyrme,brink} and the contact interactions with coupling strengths depending on the s-wave scattering length which are employed for dilute atomic gases \cite{bruun}.\\
In phenomenological effective interactions, such as Skyrme forces, a set of parameters has to be adjusted. These parameters are commonly fitted to reproduce several observables at the mean-field level (for instance, in the nuclear case, binding energies and radii of some selected nuclei and the properties of nuclear matter); this means that these interactions are constructed to be used for mean-field -based calculations. When one goes beyond the standard mean-field theories, it is not obvious at all that the same effective interactions are still suitable to be used. An additional problem arises when these interactions are contact forces: In several beyond mean-field theories, an artificial ultraviolet divergence actually appears if the range of the interaction is zero. An ultraviolet divergence already appears in the mean-field-based Hartree-Fock-Bogoliubov (HFB) \cite{RS} or Bogoliubov-de Gennes (BdG) \cite{degennes} models with zero-range forces in the pairing channel. 
Examples of beyond-mean-field models where ultraviolet divergences appear 
owing to the use of a zero-range interaction are: Higher-Tamm-Dancoff models where multiparticle-multihole configurations are introduced \cite{Quentin}, second random-phase-approximation models where not only $ph$ matrix elements are included in the matrix to diagonalize owing to the presence of two particle-two hole configurations \cite{gambacurta} and models that describe the particle-phonon coupling between individual degrees of freedom and collective vibrations \cite{be80,li07}.\\
In nuclear physics, this divergence is usually treated by introducing a momentum cutoff $\Lambda$ \cite{mog2010,mog2012} or by applying the the techniques of dimensional regularization \cite{mog2013}.\\
To summarize, when phenomenological zero--range interactions are used, two issues have to be addressed if beyond-mean-field theories are adopted:
\begin{enumerate}
\item Choosing a strategy to handle the ultraviolet divergence associated to the use of a contact force;
\item Deciding at which level the fitting procedure of the parameters has to be done and which constraints have to be included.
\end{enumerate}
In this article, we apply to a simple case a strategy to account for points 1) and 2). This strategy is quite general and can be applied to different cases and domains. We explore the possibility to fit phenomenological interactions at some beyond-mean-field level (so that these interactions are suitable to be used in beyond-mean-field models) by including the numerical momentum cutoff among the parameters of the interaction.\\
The article is organized as follows. In Sec. \ref{sec:1}, we consider symmetric nuclear matter with a simple zero-range density-dependent interaction that corresponds to the so-called $t_0-t_3$ Skyrme model in nuclear physics. The energy per particle at the second-order beyond the mean-field approximation is derived analytically. In section \ref{sec:3}, we consider nuclear matter with the full Skyrme interaction. Conclusions are summarized in Sec. \ref{sec:5}. 

\section{Second-order equation of state with the $t_0-t_3$ Skyrme model}
\label{sec:1}

In this section, we will consider symmetric nuclear matter and a zero-range density-dependent interaction written as 
\begin{eqnarray}
V(\vec{r_1}, \vec{r_2})=\delta(\vec{r_1}-\vec{r_2})\left(t_0+\frac{t_3}{6}\rho^{\alpha}\right)
\label{t0t3}
\end{eqnarray}
where $t_0$, $t_3$ and $\alpha$ are parameters. We introduce the strength $g$ of the interaction: $g(\rho)=t_0+\frac{t_3}{6}\rho^{\alpha}.$
This choice of the interaction corresponds to the so-called $t_0-t_3$ model that is a simplification of the usual Skyrme model where the non-local terms (i.e velocity dependent terms) and the spin-orbit part are neglected. Also, in nuclear physics, the finite-range Gogny force has a zero-range density dependent term.

\begin{figure}[!h]
\begin{center}
\resizebox{0.45\columnwidth}{!}{%
  \includegraphics{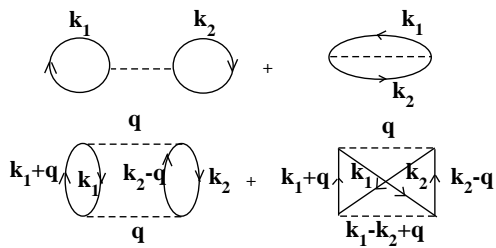} }
\caption{Direct and exchange first-order (upper line) and second-order (lower line) contributions to the total energy. }
\label{fig:1}
\end{center}       
\end{figure}
The equation of state with the simplified Skyrme interaction $V(\vec{r_1}, \vec{r_2}) $ at the mean-field level for symmetric nuclear matter is calculated by considering the first-order diagram (plus the exchange term) displayed in the first line of Fig.\ref{fig:1}:
\begin{eqnarray}
\frac{E}{A}(\rho)= \frac{3\hbar^2}{10m}
\left(\frac{3\pi^2}{2}\rho\right)^{2/3} + \frac{3}{8}t_0\rho +
\frac{1}{16}t_3 \rho^{\alpha+1}.
\label{meaneqn}
\end{eqnarray}
The second-order correction is given by the diagram (plus the exchange term) plotted in the second line of Fig \ref{fig:1}:
\begin{eqnarray}
\Delta E = d\frac{\Omega^3}{(2 \pi)^9} \int_{\vec{k_1},\vec{k_2}<k_F,|\vec{k_1}+\vec{q}|,
\vec{k_2}-\vec{q}|>k_F} d^3\vec{k_1} d^3\vec{k_2} d^3\vec{q}\;\frac{v^2(\vec{q})}{\epsilon_{\vec{k_1}} + \epsilon_{\vec{k_2}} -\epsilon_{\vec{k_1}+\vec{q}}-
\epsilon_{\vec{k_2}-\vec{q}}}
\label{eq6}
\end{eqnarray}
In Eq. \ref{eq6}, the energies $\epsilon$ are expressed as $\epsilon_{k}=\frac{\hbar^2 \vec{k}^2}{2m^*}$, where $m^*$ is the fermion effective mass. It is equal to the fermion mass $m$ in the case of the $t_0-t_3$ Skyrme model due to the absence of non-local terms.\\
$v(\vec{q})$ represents the interaction in the momentum space.\\
A simple power counting argument shows the presence of ultraviolet divergence. Whereas $k_1$ and $k_2$ are limited by the Fermi momentum $k_F$, $q$ can be arbitrarily large and for large values of this momentum the second-order correction behaves like 
\begin{eqnarray*}
\Delta E\propto \int^{\infty}d^3\vec{q}\;\frac{v^2(\vec{q})}{q^2}.
\end{eqnarray*}
In other words, the second-order correction in symmetric nuclear matter with the simplified Skyrme interaction diverges linearly with the momentum cutoff $\Lambda$ \cite{mog2010,mog2012}:
\begin{eqnarray}
\frac{\Delta E}{A}(\rho,\Lambda\rightarrow\infty)=\frac{1}{105}\left(-11+2\ln2\right)\;\chi(\rho)+\frac{\Lambda}{9k_F}\;\chi(\rho)+\mathcal{O}\left(\frac{k_F}{\Lambda}\right), \quad \chi(\rho)=-\frac{3}{4 \pi^6} \frac{m\;k_F^7}{\hbar^2 \rho}\;g^2(\rho).
\label{asym}
\end{eqnarray} 
where $k_F$ is the Fermi momentum and $A$ is the particle number.

\begin{figure}[!h]
\begin{center}
\resizebox{0.65\columnwidth}{!}{\includegraphics{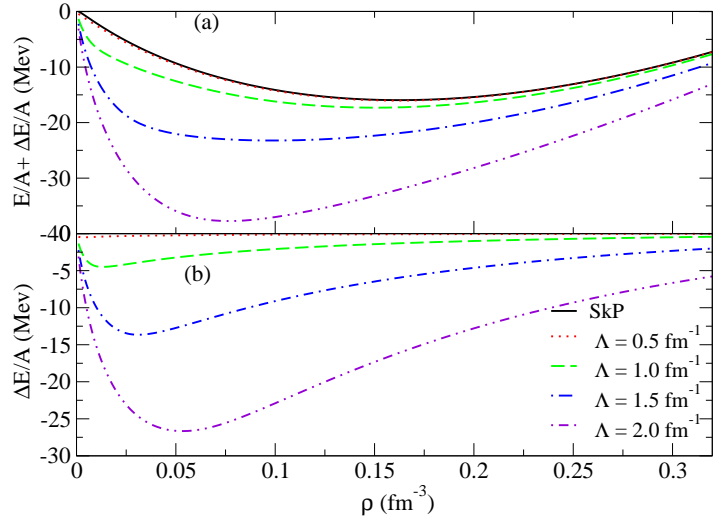} }
\caption{(a): $E/A+\Delta E/A$ as a function of the density $\rho$ (in units of fm$^{-3}$ for different values of the cutoff $\Lambda$ (in units of fm$^{-1}$) compared with the SkP-mean-field equation of state (solid black line). (b): Second-order correction $\Delta E/A$ for different values of $\Lambda$.}
\label{fig:2} 
\end{center}      
\end{figure}

In fig. \ref{fig:2}(a), $E/A+\Delta E/A$ is plotted for different values of the cutoff $\Lambda$ (from 0.5 to 2 fm$^{-1}$) and compared with the SkP-mean-field curve (solid black line). The correction $\Delta E/A$ is also plotted for different values of the cutoff $\Lambda$ in Fig. \ref{fig:2}(b). One observes that the second-order correction causes a shift of the equilibrium point to lower densities.
\vspace{0.45cm}

\begin{figure}[!h]
\begin{center}
\resizebox{0.65\columnwidth}{!}{%
\includegraphics{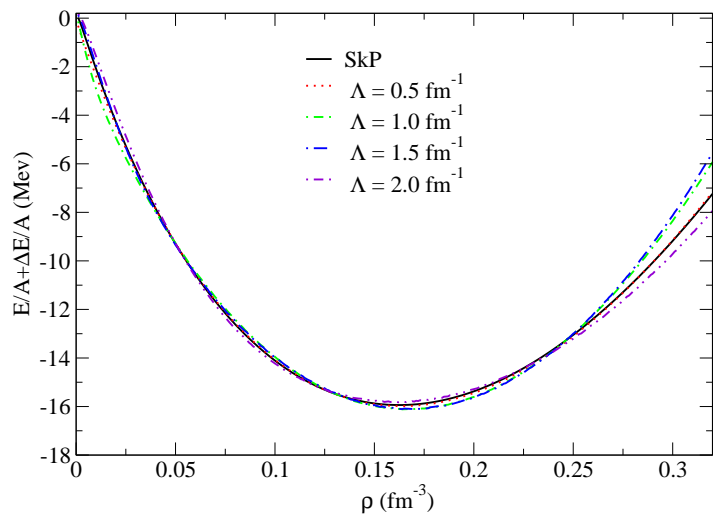} }
\caption{Refitted second-order-corrected equations of state compared with the reference SkP-mean-field equation of state.}
\label{fig:3}
\end{center}       
\end{figure}

Eq. \ref{meaneqn} actually coincides with the equation of state obtained with the SkP Skyrme parameterization \cite{skp} where no contribution coming from the velocity dependent terms $t_1$ and $t_2$ appears. In this work, the SkP-mean-field equation of state is used to make comparisons with the beyond-mean-field equations and is employed as our reference to perform the fit of the parameters.\\
For each value of the cutoff $\Lambda$, we have performed a least square fit to determine a new parameter set SkP$_\Lambda$ for the beyond-mean-field corrected equation of state $E/A+\Delta E/A$ so that the SkP mean-field EoS is reproduced. The corresponding curves obtained with the adjusted parameters are shown in Fig. \ref{fig:3} for different values of $\Lambda$. In the inset of Fig. \ref{fig:3}, the refitted EoS are plotted and compared with the SLy5 mean-field EoS (solid line). The refitted parameters are listed together with the saturation point in Table \ref{tab:1}.\vspace{0.5cm}

\begin{table}[!h]
\caption{From the second line, columns 2, 3 and 4: parameter sets obtained in the fits associated 
with different values of the cutoff $\Lambda$ 
compared with the original set SkP (first line). In the fifth column the $\chi^2/N$-value ($\chi^2$ divided by the number of fitted points) associated to 
each fit is shown. In column 6 the saturation point is shown.}
\label{tab:1}       
\begin{tabular}{ccccccc}
\hline\noalign{\smallskip}
 & $t_0$ (MeV fm$^{3}$)  & $t_3$ (MeV fm$^{3+3\alpha}$)& $\alpha$ & $\chi^2/N$  &  $\rho_0$ (fm$^{-3}$) & 
$E/A(\rho_0)$ (MeV) \\
\noalign{\smallskip}\hline\noalign{\smallskip}
        SkP &   -2931.70   &   18708.97 & 1/6   & &  0.16   &  -15.95  \\
        $\Lambda=$ 0.5 fm$^{-1}$ & -2352.900   & 15379.861  &  0.217   
& 0.00004 &  0.16     & -15.96  \\
        $\Lambda=$ 1 fm$^{-1}$ & -1155.580     &  9435.246   & 0.572  &  0.00142 & 0.17   & -16.11  \\
$\Lambda=$ 1.5 fm$^{-1}$ & -754.131    & 8278.251   &  1.011 & 0.00106 & 0.17 & -16.09  \\
$\Lambda=$ 2 fm$^{-1}$ & -632.653     &  5324.848  & 0.886 & 0.00192 & 0.16  & -15.82  \\
\noalign{\smallskip}\hline
\end{tabular}
\end{table}

\section{Second-order equation of state with the full Skyrme interaction}
\label{sec:3}
\begin{figure}[!h]
\begin{center}
\resizebox{0.65\columnwidth}{!}{%
\includegraphics{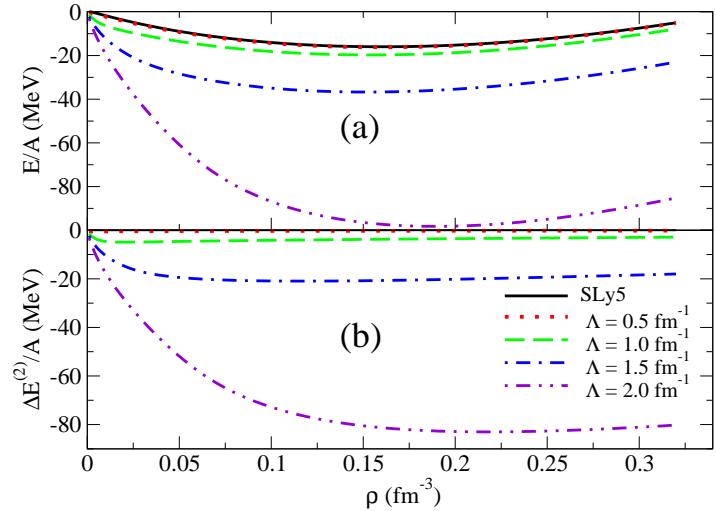} }
\caption{(a) Second-order EoS for different values of the cutoff $\Lambda$ and (b)  
second-order correction  for symmetric nuclear matter calculated with the SLy5 parameters. The SLy5 mean-field EoS is also plotted in (a) (solid line).}
\label{fig:5}
\end{center}       
\end{figure}

As it was done in section \ref{sec:1}, we treat the equation of state of nuclear matter by adding the second-order contribution to the first-order mean-field energy. In this section, we use a standard Skyrme interaction such as SLy5 in its complete form:
 \begin{eqnarray}
V({\vec r}_1, {\vec r}_2) &=& t_0(1+x_0P_{\sigma}) \delta({\vec r}_1-
{\vec r}_2)
+ \frac{t_1}{2}(1+x_1P_{\sigma})[{\vec k}'^{2}
\delta({\vec r}_1-{\vec r}_2)+
\delta({\vec r}_1-{\vec r}_2){\vec k}^{2}]+ t_2(1+x_2P_{\sigma}){\vec k}'\cdot\delta({\vec r}_1-
{\vec r}_2){\vec k}\nonumber\\&+&
\frac{t_3}{6}(1+x_3P_{\sigma})\;\rho^{\alpha}\left(\frac{\vec{r}_1+\vec{r}_2}{2}\right)
\delta({\vec r}_1-{\vec r}_2)+ iW_0 ({\sigma}_1+{\sigma}_2)\cdot[{\vec k}'\times
\delta({\vec r}_1-{\vec r}_2){\vec k}].
\label{interaction}
\end{eqnarray}
In Ref. \cite{mog2012}, the energy per particle in nuclear matter calculated at the second-order has been derived analytically using the general Skyrme force of Eq. \ref{interaction}.
Its asymptotic behaviour for large values of the momentum cutoff $\Lambda$ shows that the divergence is strong ($\sim\Lambda^5$) and that this divergence is much stronger than the linear divergence of the $t_0-t_3$ model in Eq. \ref{asym}:
\begin{eqnarray}
\frac{\Delta E^{(2)}}{A}(\delta, \rho, \Lambda \rightarrow \infty) = 
a_1(\delta,\rho)\;\Lambda^5 + a_2(\delta,\rho)\;\Lambda^3 + a_3(\delta,\rho)\;\Lambda + a_4(\delta,\rho) + O\left(\frac{k_F}{\Lambda}\right).
\label{divergence}
\end{eqnarray}
The last term $a_4(\delta,\rho)$ in Eq. \ref{divergence} has been calculated explicitly in Ref \cite{mog2013} for symmetric nuclear matter ($\delta=0$), asymmetric matter and for pure neutron matter ($\delta=1$).\\
In the upper panel of Fig. \ref{fig:5} we plot, as an example, the second-order EoS of symmetric nuclear matter obtained for different values of the cutoff $\Lambda$ (see legend), from 0.5 up to 2 fm$^{-1}$. The different equations of state are calculated by using the SLy5 Skyrme parameters and are compared with the reference mean-field SLy5 EoS (solid line in (a)). In (b) the second-order correction is plotted for the same values of the cutoff $\Lambda$. 

\begin{figure}
\begin{center}
\resizebox{0.65\columnwidth}{!}{%
\includegraphics{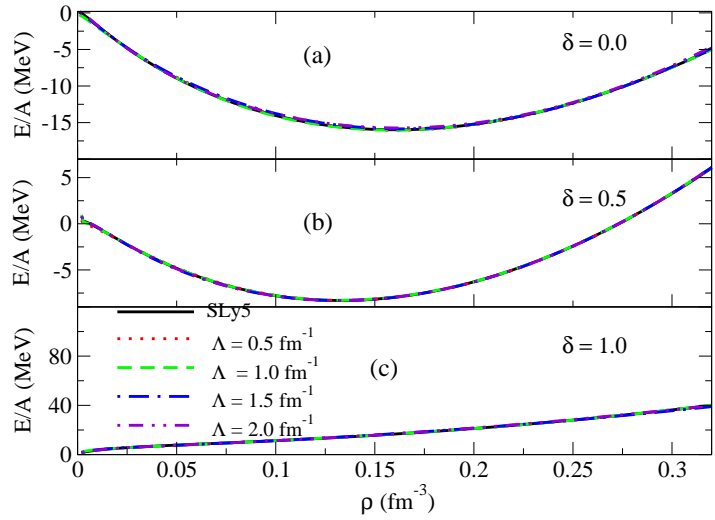} }
\caption{Refitted EoS (global fit) for symmetric (a), asymmetric $\delta=0.5$ (b), and pure neutron matter (c). The reference SLy5 mean-field curves are also plotted in the three panels (solid lines).}
\label{fig:4}
\end{center}       
\end{figure}

In Ref. \cite{mog2012}, the problem of ultraviolet divergence has been solved in the following way.
For each value of the cutoff $\Lambda$, we have performed a chi-squared minimization to determine the new sets of parameters for the beyond-mean-field corrected equation of state $E/A+\Delta E/A$. All the parameters are kept free in the adjustment procedure. The minimization has been performed on $15$ equidistant points using the following definition for the $\chi^2$:
\begin{eqnarray}
\chi^2=\frac{1}{N-1}\sum_{i=1}^{N}\left(\frac{E_i-E_{i,ref}}{\Delta E_i}\right)^2.
\end{eqnarray}
The errors or {\it adopted} standard deviations, $\Delta E_i$, are chosen equal to 1\% of the reference SLy5 mean-field energies $E_{i,ref}$. This choice is arbitrary since we are fitting a theoretical EoS where a standard deviation for this quantity has not been estimated. However, the magnitude of the $\chi^2$ defined in Eq. \ref{chi} has a clear and reasonable meaning: if it is smaller or equal to one, the reference EoS is reproduced within one standard deviation, i.e., within a 1\% average error by our second-order EoS.\\
In this work, a unique and global fit has been done to readjust the three mean-field plus second-order EoS for symmetric, asymmetric $(\delta=0.5)$ and pure neutron matter to reproduce the corresponding SLy5 mean-field curves.\\The obtained sets of parameters are presented in Table \ref{tab:2}. In Fig. \ref{fig:4} the three refitted EoS are plotted as function of the density $\rho$ for different values of the cutoff $\Lambda$. The value of the saturation density is equal to  $0.16$ fm$^{-3}$ for the four values of $\Lambda$.

\begin{table}[!htbp]
\caption{Parameter sets obtained in the fit of the EoS of symmetric, asymmetric $(\delta=0.5)$ and pure neutron matter for different values of the cutoff $\Lambda$ compared with the original set SLy5. The standard deviation, $\sigma$, estimated for the different parameters is also given. In the last columns the $\chi^2$ and the incompressibilty $K_{\infty}$ values  are shown.}
\label{tab:2}       
\scalebox{0.78}{%
\begin{tabular}{c c c c c c c c c c c c}
\hline\noalign{\smallskip}
&  $\quad t_0$ & $t_1$ & $t_2$ & $t_3$ & $x_0$ & $x_1$ & $x_2$ & $x_3$ & $\alpha$ & &$k_{\infty}$  \\
   &  $\quad \sigma_{t_0}$ & $\sigma_{t_1}$ & $\sigma_{t_2}$ & $\sigma_{t_3}$ & $\sigma_{x_0}$ & $\sigma_{x_1}$ & $\sigma_{x_2}$ & $\sigma_{x_3}$ & $\sigma_{\alpha}$ &  \\
 & (MeV fm$^3$) & (MeV fm$^5$) & (MeV fm$^5$) & (MeV fm$^{3+3\alpha}$) & & & & & & & (Mev) \\
\noalign{\smallskip}\hline\noalign{\smallskip}
SLy5 & -2484.88	&483.13     &  -549.40	&   13736.0    &   0.778	&       -0.328	&     -1.0	&   1.267&	0.16667  & &229.90\\
\hline
$\Lambda$(fm$^{-1})$ & & & & & & & & &  &$\chi^2$& \\ 
0.5 & -2022.142 &  290.312 &  1499.483 &  12334.459 &  0.481 & -5.390 & -1.304 &  0.880 &  0.259 & 0.411& 236.36 \\
    &  $ 0.49$ & $ 0.212$& $ 1.75$ & $ 4.5$ & $ 0.001173$ & $ 0.00657$ & $ 0.00020$ &  $ 0.001632$ & $ 0.000280$ &  \\
1.0 & -627.078 &  83.786 & -971.384 &  186.775  &  3.428 &  -1.252 & -1.620 & 200.360 & 0.338 & 0.540&230.52 \\
    &  $ 1.668$ & $ 0.2740$ &  $ 0.782$ & $ 0.078$ &  $ 0.00260$ & $ 0.01927$ & $ 0.00026$ & $ 0.082$ & $ 0.000314$ &  \\
1.5 & -743.227 &  112.246 & -42.816 &  5269.849 &  1.013 &  3.478 & -2.114 & 0.189 &  0.814 & 1.733&236.28 \\
    &  $ 0.306$ & $ 0.685$ &  $ 0.2972$ & $ 5.4$ &  $ 0.01415$ & $ 0.01309$ & $ 0.00519$ & $ 0.045037$ & $ 0.000784$ &  \\
2.0 & -718.397 &  573.884 & -497.766 &  6179.243 &  0.391 & -0.393 & -0.574 & 0.785  & 1.051 & 1.313&222.76   \\
    &  $ 0.343$ & $ 0.251$ &  $ 0.261$ & $ 8.33$ &  $ 0.005876$ & $ 0.001850$ & $ 0.000597$ & $ 0.017475$ & $ 0.00104$ &  \\
\noalign{\smallskip}\hline
\end{tabular}}
\end{table}
The $\chi^2$ in the global fit is composed by the three contributions and its final value is divided by three in order to make our different results comparable to one another. In other words,
\begin{eqnarray*}
\chi^2=\frac{1}{3}\left[\chi^2(\delta=0)+\chi^2(\delta=0.5)+\chi^2(\delta=1)\right].
\end{eqnarray*}
Globally, as one can see from the $\chi^2$ values, the fit is of good quality. Specifically, these values are still less than 1 up to $\Lambda=$ 1 fm$^{-1}$. Values between 1 and 2 (to be judged by considering the adopted choice of the errors in the expression of $\chi^2$) are found for larger values of the cutoff meaning that the fit is still good. Finally, we have estimated the standard deviation of the fitted parameters \cite{bevington}. This analysis allows one to asses how well the used reference data together with the adopted errors constraint the parameters of our model. In particular, the standard deviation associated to such parameters are displayed in Table \ref{tab:2}. 

\section{Conclusions}
\label{sec:5}
We have introduced a general method to handle the ultraviolet divergence generated by the use of a contact force in the evaluation of the second-order energy correction beyond the mean-field level. As a first step, we have considered symmetric nuclear matter with a simple zero-range density-dependent interaction, corresponding to the so called $t_0-t_3$ Skyrme model in nuclear physics. A cutoff regularization has been proposed to treat the ultraviolet divergence which is linear with the momentum cutoff $\Lambda$. We then proposed a fitting procedure such that the equation of state (EoS) including the second-order correction matches rather well the one obtained with the original SkP mean-field force. After that, we have extended our work to the case of nuclear matter with the nuclear Skyrme interaction. In this case, the velocity dependent terms of the Skyrme interaction have been included and the second-order equation of state shows a strong divergence ($\sim\Lambda^5$). A global fit is performed simultaneously for the EoS of symmetric, asymmetric $(\delta=0.5)$ and pure neutron matter at the second-order so that their corresponding SLy5 mean-field curves are well reproduced. These adjusted interactions display reasonable properties for nuclear matter. This opens new perspectives for future applications of this kind of interaction in beyond-mean-field models to treat finite nuclei.

\section*{Acknowledgments}
Kassem Moghrabi is thankful to Haitham Zaraket and Bira van Kolck for valuable and fruitful discussions.


\begin{thebibliography}{}

\bibitem{skyrme} T.H.R. Skyrme, Phil. Mag. 1, (1956) 1043; Nucl. Phys. 9 (1959) 615.
\bibitem{brink}  D. Vautherin and D.M. Brink, Phys. Rev. C 5, (1972) 626.
\bibitem{bruun} G. Bruun {\it et al.}, 
Eur. Phys. J. D {\bf 7}, (1999) 433.
\bibitem{RS} P. Ring and P. Schuck, {\it The Nuclear Many-Body Problem}, (Springer-Verlag Berlin Heidelberg 
New York, 1980).
\bibitem{degennes} P.G. de Gennes, {\it Superconductivity of Metals and 
Alloys}, (Perseus Books 
Publishing, L.L.C. 1966).
\bibitem{Quentin} L. Bonneau {\it et al.}, Phys. Rev. C {\bf 76}, (2007) 014304.
\bibitem{gambacurta} D. Gambacurta {\it et al.}, Phys. Rev. C {\bf 81}, (2010) 054312.
\bibitem{be80} V. Bernard and N. Van Giai, Nucl. Phys. A {\bf 348}, 75 (1980). 
\bibitem{li07} E. Litvinova {\it et al.}, Phys. Rev. C {\bf 75}, 064308 (2007).
\bibitem{mog2010} K. Moghrabi, M. Grasso, G. Col\`o, and N. van Giai, Phys. Rev. Lett. 105, (2010) 262501.
\bibitem{mog2012} K. Moghrabi, M. Grasso, X. Roca-Maza, and G. Col\`o, Phys. Rev. C 85, (2012) 044323.
\bibitem{mog2013} K. Moghrabi and M. Grasso, in preparation. 
\bibitem{skp} J. Dobaczewski {\it et al.}, Nucl. Phys. A {\bf 422}, 103 (1984).
\bibitem{bevington} P. R. Bevington and D. K. Robinson, {\it Data reduction and error analysis for physical sciences}, Second Edition (McGraw-Hill, NeW York 1992).

\end{thebibliography}
\end{document}